\documentstyle[prl,epsfig,aps,multicol]{revtex}
\newcommand{\NaCl}[1]{${\rm (NaCl)}_{#1}{\rm Cl}^-$}
\newcommand{\etal}{{\it et al.\/}} 
\DeclareSymbolFont{mathbfit}{OML}{cmm}{bx}{it}
\DeclareMathSymbol{\bfmu}{\mathord}{mathbfit}{22}
\begin{document}
\draft
\title{Structural Transitions and Global Minima of Sodium Chloride Clusters}
\author{Jonathan P.~K.~Doye}
\address{FOM Institute for Atomic and Molecular Physics, 
Kruislaan 407, 1098 SJ Amsterdam, The Netherlands}
\author{David J.~Wales}
\address{University Chemical Laboratory, Lensfield Road, Cambridge CB2 1EW, UK}
\date{\today}
\maketitle
\begin{abstract}
In recent experiments on sodium chloride clusters 
structural transitions between nanocrystals with different cuboidal shapes were detected. 
Here we determine reaction pathways between the low energy
isomers of one of these clusters, (NaCl)$_{35}$Cl$^-$.
The key process in these structural transitions is a highly cooperative 
rearrangement in which two parts of the nanocrystal slip past one another on
a $\{110\}$ 
plane in a $\langle$1\={1}0$\rangle$ direction.
In this way the nanocrystals can plastically deform, 
in contrast to the brittle behaviour of bulk sodium chloride crystals at the same
temperatures;
the nanocrystals have mechanical properties which are a unique feature of their finite size.
We also report and compare
the global potential energy minima for \NaCl{N} using two empirical potentials,
and comment on the effect of polarization.
\end{abstract}
\pacs{61.46.+w,36.40.Sx,82.30.Qt}

\begin{multicols}{2}
\section{Introduction}
There has been much interest in phase-like transitions in clusters. 
Most of this research has been directed at the finite analogue of the
solid-liquid transition~\cite{Berry88}, 
whereas the possibility of transitions between different solid forms has received
less attention~\cite{roseb92,Bartell95,Proykova97}. 
Some examples are known: 
when supported metal clusters are observed by electron microscopy, their
structure can change between icosahedral, decahedral and close-packed~\cite{Ajayan88,Marks94}.
However, the role of the electron beam and the surface is uncertain. 
Similar transitions have also been suggested for free atomic 
clusters of certain sizes, which are driven by entropy differences between 
the icosahedral, decahedral and close-packed morphologies~\cite{JD95c,JD98a,JD98d}.
However, it is not clear how these changes could be detected even if the large 
free energy barriers involved~\cite{JD98e} are surmountable.

Coexistence between solid-like isomers of small binary salt clusters has previously
been described by Rose and Berry~\cite{roseb92}.
More recently, clear examples of structural transitions have emerged from experiments on NaCl clusters.
These clusters have only one energetically favourable morphology: 
the magic numbers that appear in mass spectra correspond to cuboidal 
fragments of the bulk crystal (rocksalt) lattice~\cite{Campana81,Pflaum85,Twu90},
hence the term nanocrystals.
Indirect structural information comes from the experiments 
of Jarrold and coworkers which probe the mobility of size-selected cluster ions.
For most (NaCl)$_N$Cl$^-$ with $N>30$, 
multiple isomers were detected which were assigned as
nanocrystals with different cuboidal shapes~\cite{Dugourd97a}.
The populations in the different isomers were not initially equilibrated, but 
slowly evolved, allowing rates and activation energies for the
structural transitions between the nanocrystals to be obtained~\cite{Hudgins97a}.
Based on the small values of the activation energies, Hudgins {\it et al.\/} suggested that 
the rearrangement mechanisms
might involve a sequence of surface diffusion steps.

The aim of the present work is to examine this hypothesis by identifying the mechanisms
for one of the clusters that was studied experimentally, (NaCl)$_{35}$Cl$^-$.
We also report the global potential energy minima of (NaCl)$_N$Cl$^-$ for two different empirical
potentials which enables us to comment on the effects of including polarizabilities
and hence induction energies.
Methods are outlined in section \ref{sect:methods} and the global minima are
discussed in \ref{sect:global}.
Then in section \ref{sect:results} we describe the mechanism that we found to 
mediate the structural transitions in (NaCl)$_{35}$Cl$^-$. 
Finally, in section \ref{sect:disc} we discuss the implications
of this mechanism for the mechanical properties of NaCl nanocrystals and its 
relevance to other alkali halides.

\section{Methods}
\label{sect:methods}

\subsection{Searching the PES}

The identification of mechanisms for the structural transitions  
of \NaCl{35}
presents a considerable challenge to the theoretician since the 
half-lives of the least stable isomers are of the order of milliseconds~\cite{Hudgins97a}, 
whereas the time scales that can be probed by conventional molecular dynamics
simulation are only on the order of nanoseconds. 
The difficulty is that in molecular dynamics or Monte Carlo simulations the system spends most of the time 
vibrating about a minimum on the PES with transitions between minima 
occurring only rarely. 
One approach to enhancing the rate of occurrence of rare events, 
such as these structural transitions, is to bias the system towards the transition 
regions using umbrella sampling~\cite{Torrie};
this method is particularly suited to calculating free energy barriers.
Here we use an approach which is ideal for finding reaction pathways for complex processes. 
Previously it has been used to find 
a pathway between the face-centred cubic
global minimum and the lowest energy icosahedral minimum of 
a 38-atom Lennard-Jones cluster~\cite{JD97a}, 
and to identify relaxation mechanisms in amorphous silicon~\cite{Barkema98a}.
Steps are taken directly between minima~\cite{JD97a,Barkema96a,Mousseau97},
thus allowing large distances on the PES to be traversed.
To do this, we first find a transition state connected to the current minimum
using the eigenvector-following technique~\cite{cerjanm81,simonsjto83,onealts84,banerjeeass85,baker86,baker87}. 
We then calculate the corresponding rearrangement mechanism and thereby obtain the new minimum. 
Finally, we decide whether to accept the step to the new minimum, typically
on the basis of a Metropolis criterion~\cite{Metropolis}. 
By repeating this process the system performs a walk amongst connected minima
on the PES.

The steps between adjacent minima are not biased to go 
in any particular direction.
Nevertheless, by performing an extensive search of the low-lying minima 
we were able to find paths between the various rocksalt-type isomers of \NaCl{35}.
In the process over 4500 minima and 5500 transition states were characterized.
Although there is no guarantee that we have found the shortest or lowest barrier 
rearrangements between the nanocrystals, we are confident that our paths are good 
estimates and representative of the optimal paths.

\subsection{Potentials and geometry optimization}

Two popular empirical potentials for NaCl were considered in the present work.
The first is the Tosi-Fumi parameterization of the Coulomb plus Born-Meyer (C+BM) potential~\cite{Tosi64}: 
\begin{displaymath}
E = \sum_{i<j}\left({q_iq_j\over r_{ij}} + A_{ij}e^{-r_{ij}/\rho}\right),
\end{displaymath}
where $q_i$ is the charge on ion $i$, $r_{ij}$ is the distance between ions
$i$ and $j$ and $A_{ij}$ and $\rho$ and parameters~\cite{Tosi64}.
We have also considered the more complex potential fitted by Welch \etal~\cite{Welch75}
for which the full vector form has previously been given by Phillips \etal~\cite{phillipscb91}:
\begin{eqnarray*}
E &=& \sum_{i<j}\Big({q_iq_j\over r_{ij}} + A_{ij}e^{-r_{ij}^{\rm eff}/\rho}
    - {q_i (\bfmu_j\bullet{\bf r}_{ij})\over r_{ij}^3} 
    - {q_j (\bfmu_i\bullet{\bf r}_{ji})\over r_{ij}^3} \\
   &&  -3{ (\bfmu_j\bullet{\bf r}_{ij})
           (\bfmu_i\bullet{\bf r}_{ij})\over r_{ij}^5} 
       + {\bfmu_i\bullet\bfmu_j\over r_{ij}^3} \Big) 
       + \sum_i { \bfmu_i^2\over 2\alpha_i }, \\
\end{eqnarray*}
where 

\begin{displaymath}
r_{ij}^{\rm eff}= {\bf r}_{ij} +{\bfmu_i\over Q_i}-{\bfmu_j\over Q_j}
\end{displaymath}
and $r_{ij}^{\rm eff}=|{\bf r}^{\rm eff}|$.
(There is a small typographical error in equation (3) of reference~\cite{phillipscb91}.)
The above formulae are given in atomic units.
Adopting the notation of Stone~\cite{AJSbook}, component $\alpha$ of the
induced dipole vector at site $B$ due to site $A$, $\mu^B_\alpha$, is:
\begin{displaymath}
\mu^B_\alpha = {1\over2}\sum_{\alpha'}\alpha^B_{\alpha\alpha'} F^A_{\alpha'}({\bf B}),  
\end{displaymath}
where {\bf A} and {\bf B} are the position vectors of the respective sites.
The total induced moment in the present case is:
\begin{displaymath}
\mu^B_\alpha = {1\over2}\sum_{\alpha'}\alpha^B_{\alpha\alpha'} \sum_{A'}\left(
               \sum_\beta \mu_\beta^{A'} T^{A'B}_{\alpha'\beta} - q^{A'}T^{A'B}_{\alpha'} \right),
\end{displaymath}
where 
\begin{eqnarray*}
T^{A'B}_{\alpha'} &= &-{R_{\alpha'}\over R^3}, \\
T^{A'B}_{\alpha'\beta} &=& {3 R_{\alpha'}R_\beta-R^2\delta_{\alpha'\beta} \over R^5}, \\
{\bf R} &=& {\bf B}-{\bf A'}, \qquad {\rm and} \qquad R=|{\bf R}|.
\end{eqnarray*}
Two approaches are available for finding the dipole moments, namely iterating the
equations for $\mu^B_\alpha$ to self-consistency or rearranging the equation as
\begin{eqnarray*}
{\bf M}\bfmu = {\bf Y} \qquad {\rm so\ that} \qquad \bfmu = {\bf M}^{-1}{\bf Y},
\end{eqnarray*}
where $\bfmu = (\bfmu^1,\bfmu^2,\ldots)$.
In the present work matrix inversion was used to obtain the self-consistent
dipole moments and the first analytic derivatives of the energy.

The inclusion of polarizabilities in the Welch potential makes this functional
form much more expensive to evaluate than the simple C+BM form.
Hence we conducted the most extensive searches of the PES
with the latter potential and then reoptimized stationary points and pathways
with the Welch form. Global minima were located using a guiding function approach
for the Welch potential, as discussed in the next section.
Transition states were located using a modified eigenvector-following approach.
The basic algorithm has been described before~\cite{Wales94b,pentamer}, and
was used in the present work with numerical second derivatives for the Welch potential. 
We also employed a new approach which does not require second derivatives and
is more efficient~\cite{newEF}.

\section{Global Optimization}
\label{sect:global}

Global potential energy minima were located for \NaCl{N} up to $N=35$ for both
of the empirical potentials described above. We employed the `basin-hopping'
or Monte Carlo minimization\cite{Li87a} technique which has recently
been investigated for a variety of atomic and molecular 
clusters\cite{JD98a,gmin,morse2,SCgmin,gminH2O,NeDIM,ArClDIM}.

\end{multicols}
\begin{figure}[p]
\begin{center}
\epsfig{figure=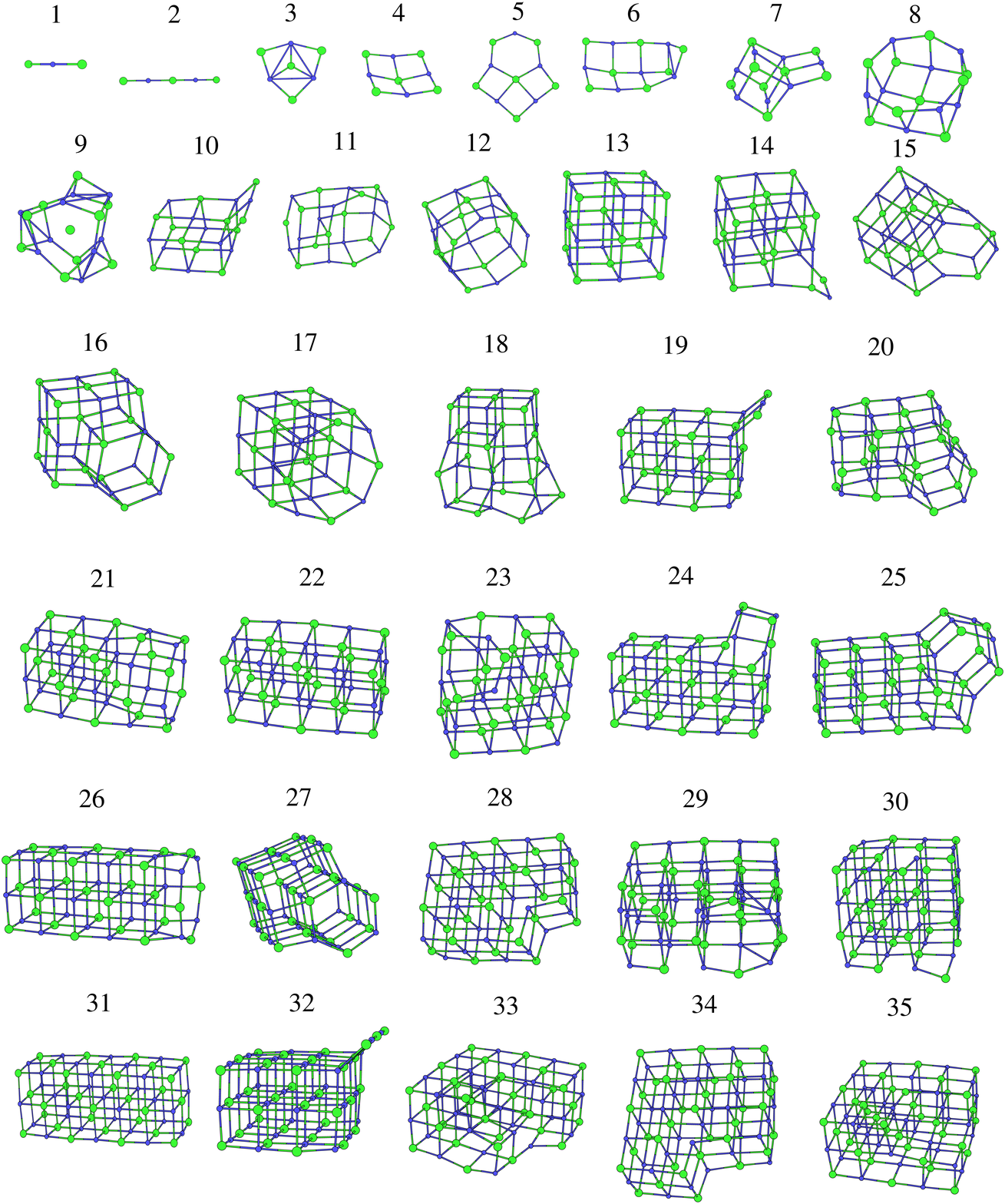,width=18cm}
\vglue-0.2cm
\begin{minipage}{18cm}
\caption{\label{fig:C+BM}Global minima for the Coulomb plus Born-Meyer potential.
Lines have been drawn using a distance cut-off to guide the eye.
The sodium ions are represented by the smaller, darker circles and 
the chloride ions by the larger, lighter circles. }
\end{minipage}
\end{center}
\end{figure}

\begin{figure}[p]
\begin{center}
\epsfig{figure=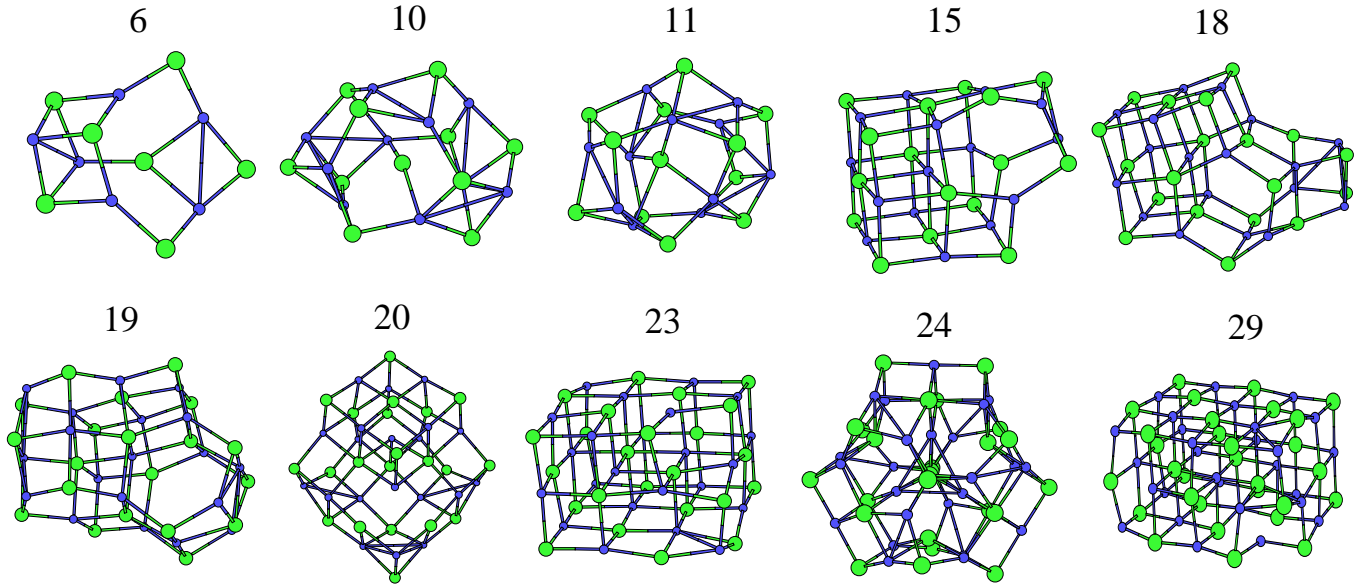,width=18cm}
\vglue-0.2cm
\begin{minipage}{18cm}
\caption{\label{fig:Welch}Global minima for the Welch potential
corresponding to local rather than global minima of the Coulomb
plus Born-Meyer potential.
Lines have been drawn using a distance cut-off to guide the eye.
The sodium ions are represented by the smaller, darker circles and 
the chloride ions by the larger, lighter circles.}
\end{minipage}
\end{center}
\end{figure}
\begin{multicols}{2}

Canonical Monte Carlo (MC) sampling was used to explore the transformed PES,
as described elsewhere~\cite{JD98a,gmin}.
For the C+BM potential five runs
of 5000 MC steps were performed for every cluster size with each run starting from a random point.
Two short runs of 200 steps each were also initiated using the lowest
minima for \NaCl{N-1} and \NaCl{N+1} as seeds.
The maximum step size for the displacement of any Cartesian coordinate
was dynamically adjusted to give an acceptance ratio of 0.5 for a temperature
corresponding to 316\,K (474\,K
for the larger clusters).
Final values for the maximum displacement were typically around 1.5 \AA.
To restrict the configuration space to bound clusters
we reset the coordinates to those of the current minimum in the Markov chain at each step~\cite{whitem98}.

Since the Coulomb potential is long-ranged these clusters actually represent
a rather easy global optimization problem because there are fewer minima
on their potential energy surfaces~\cite{JD95c,morse2,doyew96}.
Hence we are confident that most of the lowest energy structures we have
identified for the simpler potential are the true global minima.
In fact, the lowest minimum was usually the same for each of the five runs at
every size considered. The Welch potential is much more time consuming to
evaluate, and in this case only three runs of 4000 steps each were performed.
We also employed a guiding function technique, as suggested by Hartke~\cite{hartke96},
where the simpler C+BM was used for partial geometry optimization 
followed by relaxation with the full Welch potential in every quench. Once again the same
lowest energy structures were usually found in each of the three runs for every size.
The results also agree with calculations performed without a guiding function
up to $N=17$, and with 
those of Phillips \etal~\cite{phillipscb91}
up to $N=14$, although our energies are systematically lower than theirs, presumably
due to different unit conversion factors.
Our energies and geometries are all well converged with the root-mean-square force
reduced to less than $10^{-8}\,$hartree/bohr for every minimum.
The most difficult case in this size range appears to be $N=31$ where
the $7\times3\times3$ rocksalt global minimum was only found in two out of three runs for
the Welch potential. 

All the results will be provided in a downloadable format from the Cambridge
Cluster Database~\cite{Web}. The energies of the lowest minima found
for both the C+BM and Welch potentials are given in Table \ref{table:global}
and the structures are illustrated in Figures \ref{fig:C+BM} and \ref{fig:Welch}.
The lowest ten or so local minima found for the B+CM potential were also
relaxed separately under the Welch potential to establish the correspondence
between local minima. 

For most sizes in the present range the structure of the global potential minimum
is the same for the two potentials. However, the global minimum of one
potential is only a local minimum for the other potential at $N=6,\ 10,\ 11,\ 15,\
18-20,\ 23,\ 24$ and 29. For $N=2$ the global minimum is linear for C+BM
but somewhat bent for Welch. 
For $N=5$ the Welch global minimum has lower symmetry than the C+BM structure;
the apparent $D_{5d}$ symmetry is actually slightly broken on close inspection. 
The similarity of the results for the two potentials shows that the C+BM form, 
despite its simplicity, can give a reasonably reliable guide to the structure of 
these clusters and provides justification for our use of the C+BM potential
to perform surveys of the energy landscape.

\begin{figure}
\begin{center}
\epsfig{figure=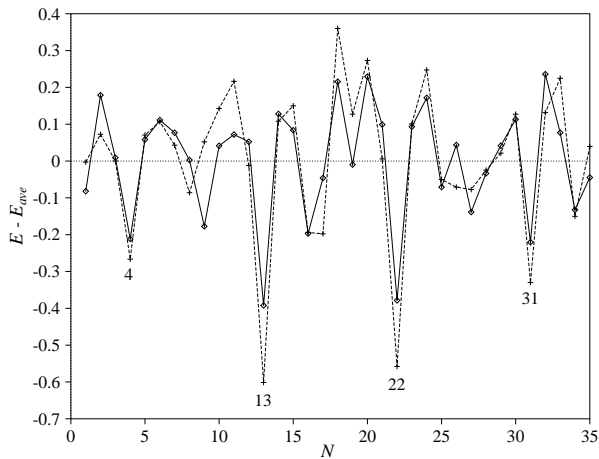,width=8.4cm}
\vglue0.1cm
\begin{minipage}{8.5cm}
\caption{\label{fig:energies}
The energies of the global minima as a function of size for the Welch (solid line)
and C+BM (dashed line) potentials.
To make the size dependence more clear, we have used a function, 
$E_{\rm ave}=a+bN^{1/3}+cN^{2/3}+dN$, as the energy zero. 
The coefficients of $E_{\rm ave}$ have been chosen to give the best fit to the 
energies. 
For the Welch potential $a=-0.5014$, $b=0.2139$, $c=0.0926$ and $d=-7.6829$.
For the C+BM potential $a=-1.6782$, $b=1.8878$, $c=-0.2828$ and $d=-7.6360$.
}
\end{minipage}
\end{center}
\end{figure}

As expected, the lowest energy structures have predominantly rock-salt structures,
and the particularly stable sizes occur when complete cuboids can be formed 
(Figure \ref{fig:energies}). 
These sizes ($N=4,13,22,31$) agree with the magic numbers observed in the
mass spectral abundance distributions\cite{Twu90}.
It is interesting to note that at a number of sizes when complete cuboids
cannot be formed a column of hexagonal rings appears
(e.g. $N=16,\ 20,\ 25,\ 27$).
One of the more unusual structures is the (NaCl)$_{24}$Cl$^-$
global minimum for the Welch potential.
It has threefold symmetry with a trigonal bipyramid in the middle containing 
three (equatorial) Na$^+$ ions and two (axial) Cl$^-$ ions.

\section{Rearrangements of \NaCl{35}}
\label{sect:results}

\begin{figure}
\vglue-2.0cm
\begin{center}
\epsfig{figure=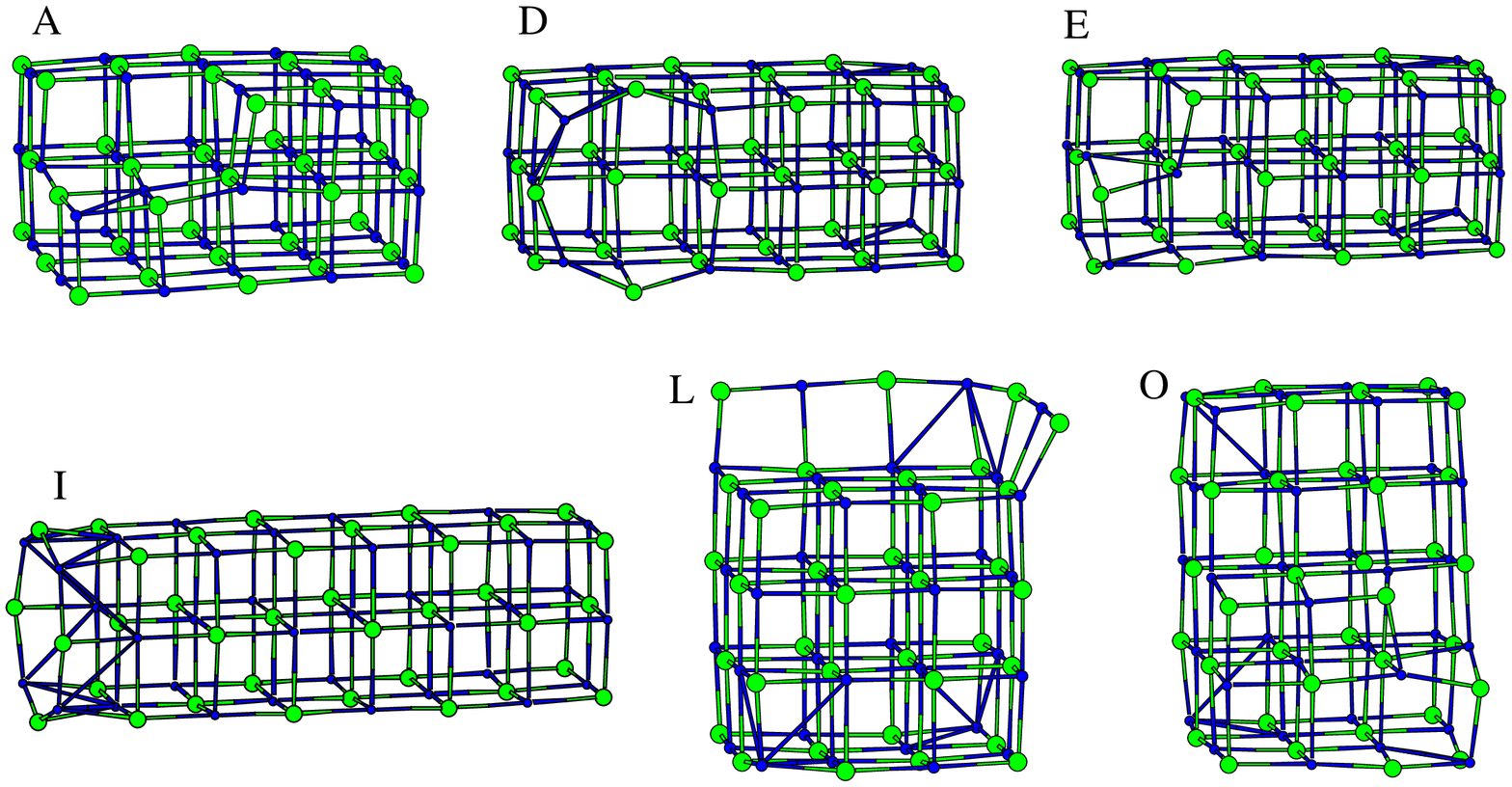,width=13.7cm}
\vglue-2.5cm
\begin{minipage}{18cm}
\caption{\label{fig:minima}
The lowest energy \NaCl{35} minima of the four types of nanocrystal for the 
Coulomb plus Born-Meyer (A, D, I and L) and Welch (A, E, I and O) potentials.
D and E differ only in the position of the vacancy at an Na$^+$ site.
The sodium ions are represented by the smaller, darker circles and 
the chloride ions by the larger, lighter circles. 
The minima are labelled as in Table \ref{table:min}.}
\end{minipage}
\end{center}
\end{figure}

In the experiments on (NaCl)$_{35}$Cl$^-$ three peaks were resolved 
in the arrival time distribution. 
These peaks were assigned on the basis of calculated mobilities 
as an incomplete $5\times 5\times 3$ cuboid, 
an incomplete $5\times 4\times 4$ cuboid, and an $8\times 3\times 3$ 
cuboid with a single vacancy~\cite{Dugourd97a,Hudgins97a}. 
However, the lowest energy minima that we found for this cluster could be divided into 
four types (Table \ref{table:min} and Figure \ref{fig:minima}), namely 
the above three nanocrystals and a $6\times 4\times 3$ cuboid with a single vacancy.
As the latter structure was not considered when the structural assignments were made,
its mobility has now been calculated using the exact hard-spheres
scattering model~\cite{Shvartsburg96,Alexprivate}.

With these new data better agreement between the calculated and observed mobilities 
can be obtained by assigning the three experimental peaks to the $6\times 4\times 3$, 
$5\times 5\times 3$ and $8\times 3\times 3$ nanocrystals in 
order of increasing drift time~\cite{Alexprivate}.
Furthermore, this reassignment gives better agreement with the energetics. 
In the experiment the clusters convert to (what we now assign as)
the $5\times 5\times 3$ nanocrystal as time progresses, indicating that this structure 
has the lowest free energy.
In our calculations the global potential energy minimum is also
a $5\times 5\times 3$ isomer. 
Moreover, the $6\times 4\times 3$ nanocrystal is second lowest in energy; it 
is 0.06$\,$eV (0.10$\,$eV for the Welch potential) above the global minimum, whereas  
the corresponding value for the lowest energy $5\times 4\times 4$ isomer is $0.21\,$eV ($0.25\,$eV).

\begin{figure}
\begin{center}
\epsfig{figure=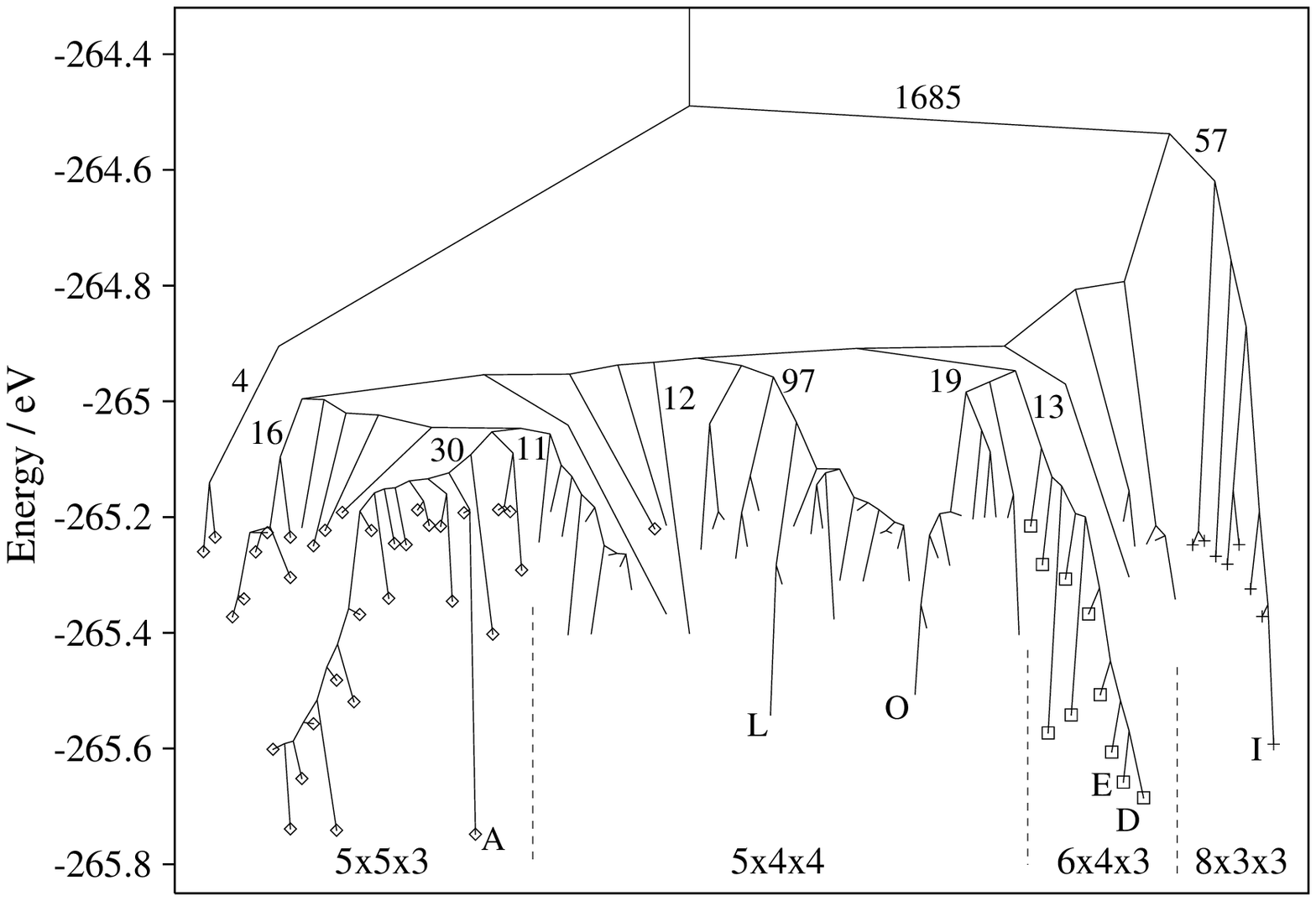,width=13.7cm}
\vglue0.1cm
\begin{minipage}{18cm}
\caption{\label{fig:funnel}
Disconnectivity graph showing the multiple-funnel character of the (NaCl)$_{35}$Cl$^-$ PES.
The branches that end at the 100 lowest energy minima are shown. 
They are marked by the type of structure: $5 \times 5 \times 3$: diamonds,
$6 \times 4 \times 3$: squares, $8 \times 3 \times 3$: crosses, and
$5 \times 4 \times 4$: unlabelled. 
The dashed lines approximately divide the tree into the different types of nanocrystal. 
The numbers denote the number of minima in a branch.
We also label the six minima that appear in Figure \ref{fig:minima} with the appropriate letter.
The energies are for the Coulomb plus Born-Meyer potential.}
\end{minipage}
\end{center}
\end{figure}

The disconnectivity graph (or tree)~\cite{BandK97,WalesMW98} in Figure \ref{fig:funnel} 
provides a graphical representation of the PES. 
It shows which minima are connected by pathways below any given total energy.
The end of each line represents a minimum on the PES and each node occurs at the energy
of the transition state which first connects the two (sets of) minima.
The tree separates the minima according to the particular rocksalt cuboid quite cleanly.
This result shows that the barriers between minima with the same basic cuboidal
shape are generally smaller than those between the different types of nanocrystal. 
This separation holds least well for the $5\times 4\times 4$ nanocrystals
because of the many different ways that the nine vacant sites can be arranged.
For example, the two lowest energy $5\times 4\times 4$ minima, L and O, have very different 
structures (Figure \ref{fig:minima}) and the energy at which the set of minima associated with O
becomes connected to the $6\times 4\times 3$ minima is lower than the energy at which 
it becomes connected to the set of minima associated with L.
Minimum O is also much closer in configuration space to the lowest energy $6\times 4\times 3$ minimum
than it is to minimum L (Table \ref{table:paths}).
Another example is provided by the two $5\times 5\times 3$ isomers on the left of the figure 
which are separated by a very large barrier from the rest of the $5\times 5\times 3$
structures because the ions occupy the opposite lattice sites to the other minima
(i.e.\ the sodium ions, not the chloride ions, are located at the corners of the nanocrystal).

One helpful way to characterize the topography of an energy landscape that 
has come from the protein folding community is in terms of funnels~\cite{Leopold,Bryngel95}.
A funnel is a set of pathways that converge to a low energy minimum.
It has been suggested that a single deep funnel underlies the ability of proteins 
to fold to a unique native structure.
Figure \ref{fig:funnel} shows that on the (NaCl)$_{35}$Cl$^-$ PES there are separate funnels 
corresponding to the $5\times 5\times 3$, $6\times 4\times 3$ and $8\times 3\times 3$ nanocrystals and
a number of small funnels associated with the $5\times 4\times 4$ nanocrystal.

\end{multicols}

\begin{multicols}{2}

\begin{figure}
\begin{center}
\epsfig{figure=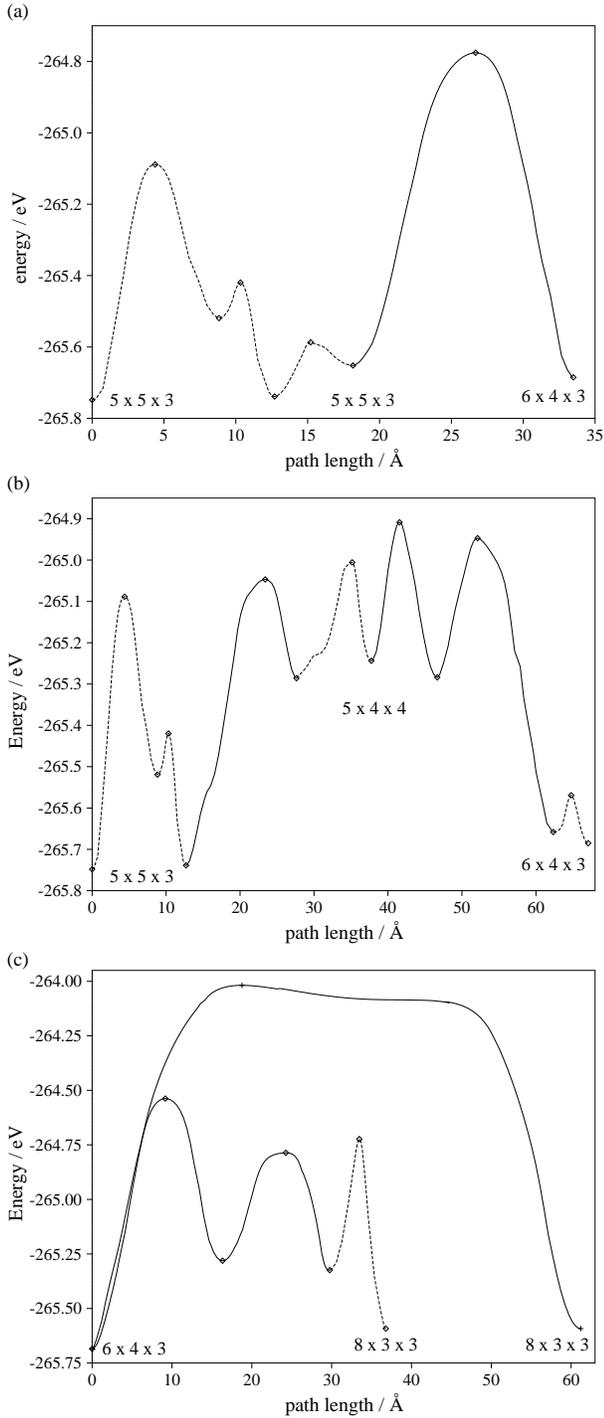,width=8.2cm}
\vglue 0.1cm
\begin{minipage}{8.5cm}
\caption{\label{fig:epaths}Energy profiles of pathways between different \NaCl{35} isomers.
(a) The shortest path and (b) the lowest energy path between the 
lowest energy $5 \times 5 \times 3$ and $6 \times 4 \times 3$ isomers.
(c) Two pathways between the lowest energy $6 \times 4 \times 3$
and $8 \times 3 \times 3$ isomers;
the shorter pathway is the pathway with the lowest barrier.
The sections of the profile with solid lines are illustrated
in Figures \ref{fig:paths} and \ref{fig:path.2}. 
The energies are for the Coulomb plus Born-Meyer potential.}
\end{minipage}
\end{center}
\end{figure}

This characterization of the PES helps to explain 
the initial presence of the metastable isomers in the experiments 
by Jarrold and coworkers.\cite{Dugourd97a,Hudgins97a}
The energy landscape 
efficiently `funnels' high energy structures into rocksalt-type minima~\cite{Ball,Rose93b}. 
However, the barriers between the different funnels are large, leading
to a separation of time scales for relaxation 
down the PES to a rocksalt structure and conversion
of the metastable nanocrystals to the global minimum.
In Ref.\ \onlinecite{JD96c} we predicted that this behaviour would occur for 
alkali halide clusters.

\begin{figure}
\begin{center}
\epsfig{figure=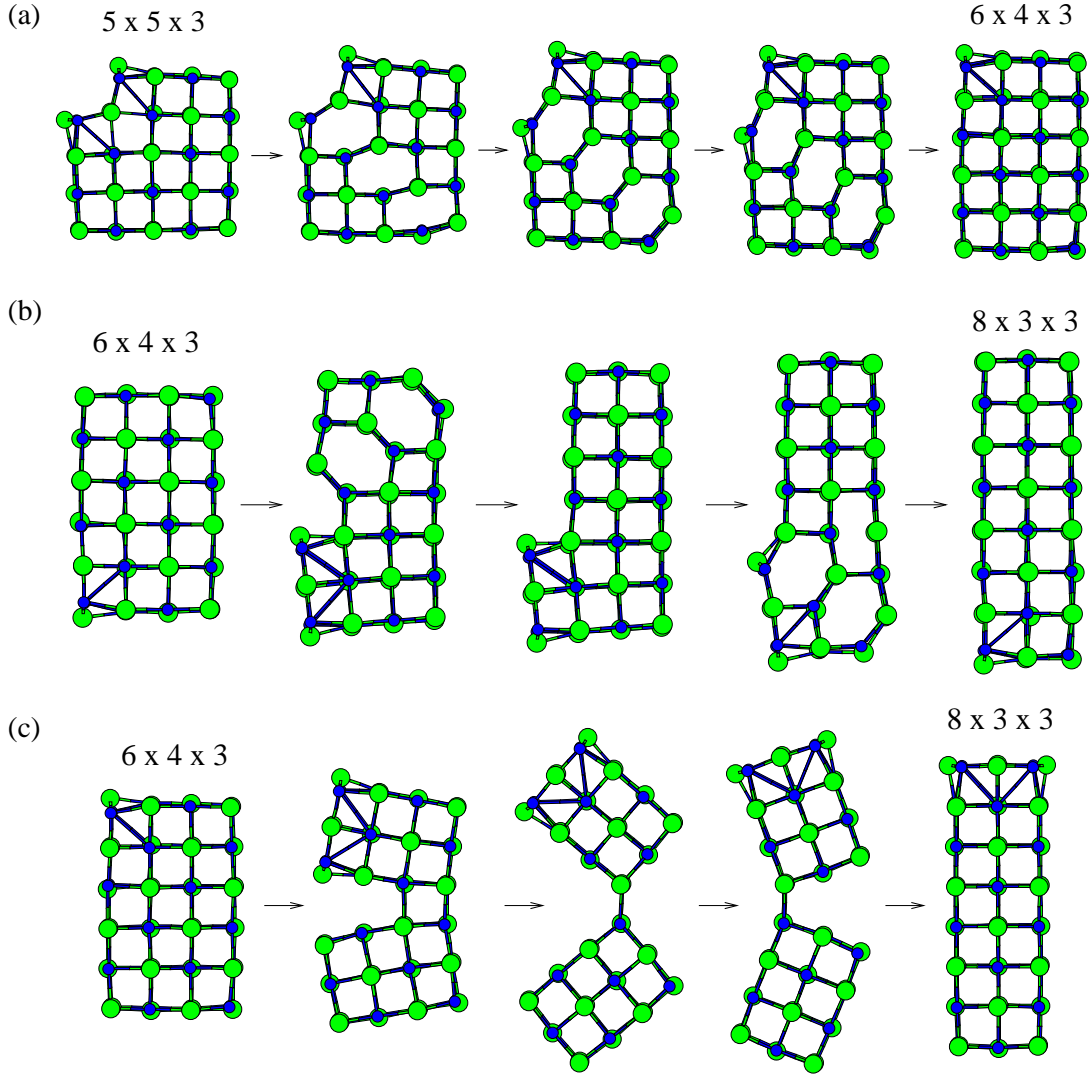,width=16cm}
\vglue-0.2cm
\begin{minipage}{18cm}
\caption{\label{fig:paths}Portions of the three rearrangement mechanisms corresponding to Fig.~\ref{fig:epaths}a and c.
(a) A glide rearrangement which converts a $5\times 5\times 3$ into a $6\times 4\times 3$ isomer.
(b) Two glide rearrangements which are part of the lowest energy path between the
$6\times 4\times 3$ and $8\times 3\times 3$ isomers.
(c) A higher energy `hinge' mechanism which converts a $6\times 4\times 3$ isomer
directly to an $8\times 3\times 3$ isomer. 
}
\end{minipage}
\end{center}
\end{figure}

\begin{figure}
\begin{center}
\epsfig{figure=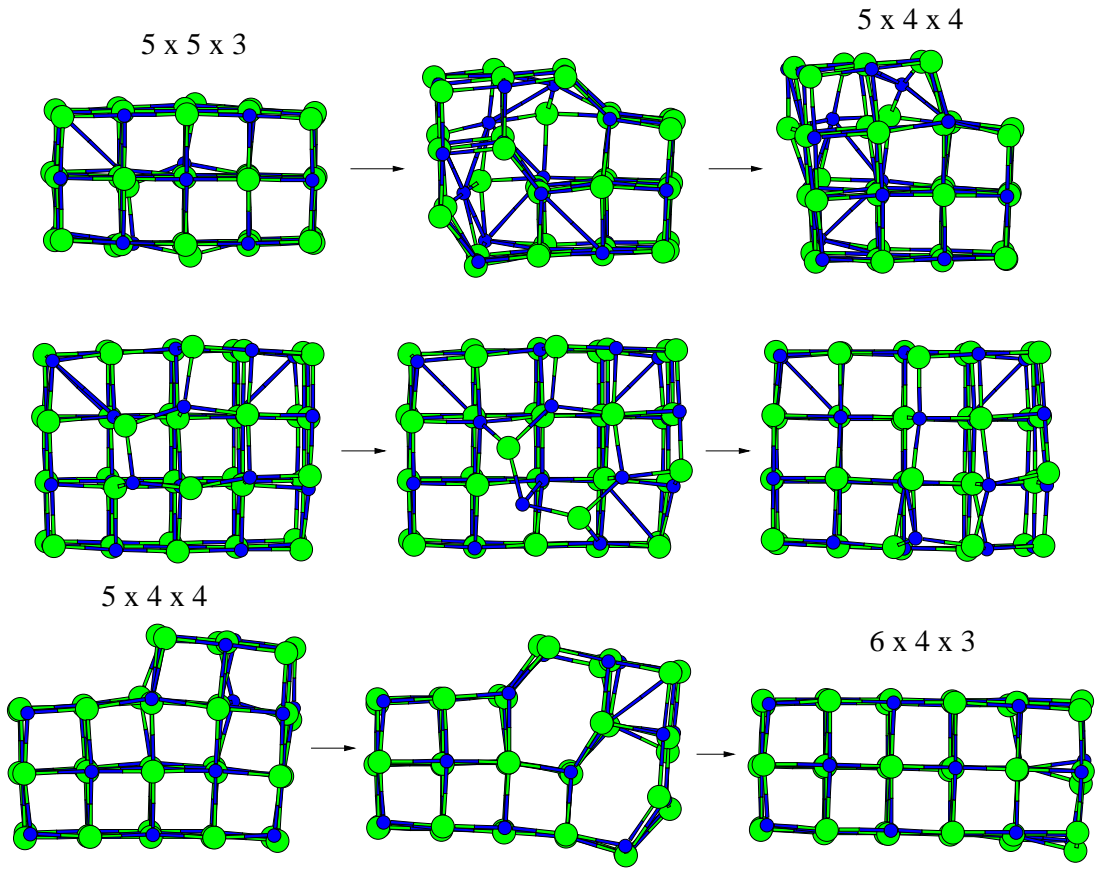,width=15cm}
\vglue-0.6cm
\begin{minipage}{18cm}
\caption{\label{fig:path.2} Three rearrangements from the lowest energy pathway between 
the lowest energy $5\times 5\times 3$ and $6\times 4\times 3$ minima, for which the energy profile
is shown in Fig.~\ref{fig:epaths}b. 
The middle frame is the transition state and on either side are the two minima.}
\end{minipage}
\end{center}
\end{figure}

To understand the mechanisms for the structural transitions we have visualized many
pathways connecting nanocrystals with different dimensions. 
In virtually all the pathways the major shape changes are achieved by 
the same type of mechanism. 
A typical example of this process occurs in the shortest path between 
the lowest energy $5\times 5\times 3$ and $6\times 4\times 3$ isomers.
The path involves four sequential transition states (Fig.~\ref{fig:epaths}a), 
the first three of which mediate local rearrangements of the empty sites in the 
$5\times 5\times 3$ cuboid. 
The fourth rearrangement causes the major shape change and
is depicted in Fig.~\ref{fig:paths}a. 
In this `glide' mechanism the two halves of the cluster slide past one another on 
a $\{110\}$ plane in a $\langle$1\={1}0$\rangle$ direction. 
Other examples of this mechanism are illustrated in Figures \ref{fig:paths}b and \ref{fig:path.2}.

Although the path shown in Figure \ref{fig:epaths}a and \ref{fig:paths}a
is the shortest between the lowest energy $5\times 5\times 3$ and $6\times 4\times 3$
minima it is not the lowest in energy.
The latter pathway (Figure \ref{fig:epaths}b and \ref{fig:path.2}) is more complicated;
rather than passing directly between the two nanocrystals it goes via 
a number of $5\times 4\times 4$ minima, some of which are structurally similar to minimum O.
Again the glide mechanism mediates the main structural changes between a $5\times 5\times 3$
and a $5\times 4\times 4$ nanocrystal (top of Figure \ref{fig:path.2}) and from a $5\times 4\times 4$ 
to a $6\times 4\times 3$ nanocrystal (bottom of Figure \ref{fig:path.2}).
The former mechanism is not quite so clear-cut since the glide is 
coupled with local defect motion on the far side of the crystal.
The middle rearrangement in Figure \ref{fig:path.2} actually involves 
the highest energy transition state on the pathway. 
In this rearrangement a triangle of ions slides over the surface of the nanocrystal.
The final rearrangement in the pathway (not depicted in Figure \ref{fig:path.2})
is the conversion of minimum E to D by the motion of a single sodium ion.

The lowest energy pathways to the $8 \times 3 \times 3$ nanocrystal all
occur via the same transition state 
The pathway from the $6 \times 4 \times 3$ nanocrystal
is illustrated in Figures \ref{fig:epaths}c and \ref{fig:paths}b.
Two glide rearrangements convert the lowest energy $6 \times 4 \times 3$ isomer
into an $8 \times 3 \times 3$ isomer. 
In the final rearrangement on this pathway a single sodium ion moves so that the vacancy in the 
$8 \times 3 \times 3$ nanocrystal occupies the site of lowest energy.

It is not hard to see why the glide mechanism is favourable.
On either side of the $\{110\}$ plane are rows of oppositely-charged ions (these can be clearly
seen on the top face of the nanocrystals in Fig.~\ref{fig:paths}a and b). 
When the nanocrystal slides on this plane in the direction of these rows, no ion
comes any closer to the nearest ion of the same charge. 
This situation only holds for $\{110\}$ planes and the $\left<1\overline{1}0\right>$ direction.
The activation barrier arises from the loss of
some favourable contacts between oppositely-charged ions. 
It is also interesting to note that this plane and direction correspond to the
primary slip system for dislocation motion in bulk NaCl crystals~\cite{Sprackling}.

One other mechanism by which major shape changes can occur 
is illustrated in Figure \ref{fig:paths}c.
In this `hinge' process the two halves of the cluster split apart and rotate about a common
edge until they meet up again.
Although spectacular this mechanism is unlikely to be relevant to the experiments
because the barrier is probably too large (Fig.~\ref{fig:epaths}b) 
to be surmountable at the appropriate temperature; the transition state is 
1.57$\,$eV (1.21$\,$eV for the Welch potential) above the $8 \times 3 \times 3$ minimum.

\end{multicols}
\begin{multicols}{2}

From the experiments of Jarrold and coworkers rates and activation energies were obtained 
for the conversion of the metastable nanocrystals to the global minimum~\cite{Hudgins97a}. 
Therefore, it would be useful if we could make estimates of the activation energies 
to provide independent support for our results.
However, this is non-trivial. 
There are a number of possible approaches.
For a complex process one needs to consider not the energy barrier, but the free energy barrier.
The latter barrier could be computed by umbrella sampling~\cite{Torrie} if a suitable order parameter 
was found which could distinguish the different nanocrystals. 
The rate of barrier crossing could then be calculated by performing a series of simulations 
which start from the top of the barrier~\cite{Chandler,Ruiz97}.
A second approach is to use a master equation to model the flow of 
probability between the minima on the PES~\cite{Kunz95}.
However, to achieve reasonable results one would need a larger set of minima and 
transition states than we have obtained here. 
Furthermore, this approach also requires an expression for the rate constants for transitions between minima. 
Although these are easily calculated within the harmonic approximation using RRKM theory, 
more accurate rate constants which include the effects of anharmonicity are far harder to obtain.
There are also two recently-developed techniques which could possibly be applied to 
calculate the rate constants: the macrostate variational method\cite{Ulitsky98a} and
the transition path sampling approach of Chandler and coworkers\cite{Dellago98a,Dellago98b}.

All these approaches are computationally demanding and beyond the scope of the present work. 
However, we can obtain an estimate of the activation energy if we make the approximation that 
the only important path between the nanocrystals is the one with lowest energy.
The activation energy can then be equated with the energy difference between 
the highest energy transition state on this path and the starting minimum.
At absolute zero this energy barrier is equal to the free energy barrier,
however at non-zero temperatures the free energy barrier is likely to be
reduced by the entropy associated with the multiplicity of paths between the two states. 
This interplay of energy and entropy has been observed in the free energy 
barrier between the lowest energy face-centred-cubic and icosahedral 
minima for a 38-atom Lennard-Jones cluster~\cite{JD98e}.

\end{multicols}
\begin{multicols}{2}

In Table \ref{table:paths} we give our estimates for the activation energies of pathways
between the lowest energy isomers of each cuboidal nanocrystal.
In the experiment it was only possible to find the activation energies for rearrangements
between the $6\times 4\times 3$ and $8\times 3\times 3$ nanocrystals 
and the $5\times 5\times 3$ nanocrystal; 
the values were $0.57\pm 0.05\,$eV and $0.53\pm 0.05\,$eV, respectively~\cite{Hudgins97a}. 
These values are of the same order of magnitude as our estimates of the activation energies 
(Table \ref{table:paths}), although somewhat lower than we find for these 
specific transformations. 
The differences could be due to the empirical potentials employed or
the approximations involved in our estimation of the activation energy,
which would probably produce overestimates, as observed.
It is also possible that we may not have found the lowest energy pathways. 
Furthermore, the barriers are calculated with respect to the lowest energy minima
for a particular type of nanocrystal. 
However, as the barriers between minima with the same cuboidal shape 
can be quite large (e.g. in Figure \ref{fig:epaths}a there is a barrier between two
$5\times 5\times 3$ isomers which is 0.66eV) it is not certain that the nanocrystals in the 
experiment are in the lowest energy minimum for that shape.

Given these uncertainties it is not yet possible to use comparison of the experimental and
estimated activation energies to provide additional confirmation of the glide mechanism.
However, the universality of the glide mechanism amongst the many low energy paths that 
we have discovered  strongly suggests that this process does mediate the structural transitions.

Although most of the results in this section are for the C+BM potential
those we have obtained using the Welch potential encourage us to
believe that the C+BM form provides a reasonable description of the interactions.
The global minimum for \NaCl{35} is the same for the two potentials.
Furthermore, the low energy minima are similar (Table \ref{table:min}). 
There are a few changes in the energetic ordering of the minima; most significantly
the $5\times 4\times 4$ minimum O is significantly lower than minimum L
for the Welch potential.
It is also interesting to note that the estimates of the activation energy are systematically
lower when the Welch potential is used (Table \ref{table:paths}) bringing the values
closer to those found in experiment\cite{Hudgins97a}.

\section{Discussion}
\label{sect:disc}

At the transition state of the glide mechanism 
there are stacks of hexagonal rings perpendicular to the direction of sliding (Figure \ref{fig:paths}). 
Hexagonal rings are a common motif in small alkali halide clusters~\cite{Phillips91}, 
and when the ratio of the cation to anion size is smaller than for NaCl,
e.g. in the lithium halides, NaBr and NaI, the hexagonal structures are more stable than for NaCl~\cite{Aguado}. 
Hence the glide mechanism might be even more favourable in these systems.
To prove this supposition calculations similar to those in this paper could be performed. 
Furthermore, if this suggestion is correct the dependence of the kinetics of structural transformation
on the particular alkali halide could be then used as an experimental test of the 
mechanism we have found here.

As the glide mechanism is a cooperative mechanism, unlike the surface diffusion mechanism 
originally suggested\cite{Hudgins97a}, it is expected that the barriers for this type of
rearrangement would generally increase with size (although as with any cluster property it is likely
that specific size effects would also be superimposed on this trend). 
The structural transitions would become more difficult with increasing size.
Again this could probably be used as an experimental test of the mechanism.

The glide mechanism also has implications for the mechanical properties
of these nanocrystals. 
It allows them to deform spontaneously and plastically to 
a thermodynamically more stable structure; in other words, the nanocrystals are soft.
Similar homogeneous slip mechanisms, where whole planes of atoms slide past one another, 
have been seen in simulations of strained metal nanowires~\cite{Finbow97,Sorensen98}.
As the barrier to these cooperative processes increases with the area of the sliding surface,
they become less feasible as the size of the system increases.
For the metal nanowires the increased barrier leads to a change in 
the slip mechanism to a more localized process, 
namely dislocation motion, with increasing size~\cite{Sorensen98}; 
the dislocations maintain the ductility of the metal nanowire.
In contrast, for NaCl nanocrystals, at temperatures relevant to the experiments considered
here (7-67$^\circ$C~\cite{Dugourd97a}), dislocation motion 
is much more difficult than for metals.
Therefore, the increasing barrier appears to lead to a dramatic change in mechanical properties. 
The nanocrystals become harder as their size increases, 
until a point is reached where an applied stress is more likely to cause fracture 
than plastic deformation, and the typical brittle behaviour of bulk NaCl crystals is recovered.

The mechanical properties of these NaCl nanocrystals provide 
another example of the unique finite-size properties of clusters.
It might be possible to confirm the increased plasticity we predict
using a microscope tip to deform NaCl nanocrystals soft-landed on a surface.

Previously, Ball \etal\
have compared relaxation to the global minimum of $\rm (KCl)_{32}$ 
with Ar$_{19}$ in terms of `monotonic sequences'~\cite{Ball}.
On this basis the salt cluster was described as a `structure-seeker' because 
rapid removal of kinetic energy usually leads to rocksalt structures; in
contrast it is relatively easy to quench Ar$_{19}$ into a defective double
icosahedron. The present results for \NaCl{35} are relevant to this relation
between the energy landscape and relaxation dynamics, 
regardless of whether the glide mechanism is operative in $\rm (KCl)_{32}$.

Assuming that the model
$\rm (KCl)_{32}$ cluster discussed elsewhere has a landscape similar to
\NaCl{35} we would expect different cuboidal rocksalt morphologies to lie
at the bottom of separate funnels. Relaxation within each funnel is efficient,
but relaxation between funnels will probably occur on a much longer time scale.
Hence we can call the salt clusters `structure-seekers' if by `structure' we
mean any of the cuboidal minima. By the same token, then, we should perhaps
have bunched the defective double icosahedra together with the global minimum
for Ar$_{19}$ in the previous study~\cite{Ball}. In fact Ar$_{19}$ is a `magic number'
cluster with an essentially single funnel landscape
and so Ar$_{19}$ is also quite an efficient `structure-seeker'.
A better contrast would be provided by a cluster bound by a short range potential,
where the landscape would be rougher, or by Ar$_{38}$, whose double funnel
landscape we have investigated elsewhere~\cite{JD98a,JD98d,JD98e,JD97a,WalesMW98}.
The point is that the salt cluster relaxes faster down any one funnel than
Ar$_{19}$, but may not reach the global minimum except over a longer time scale.

\acknowledgements

We would like to thank Alex Shvartsburg and Martin Jarrold for calculating the 
mobilities of some of the structures and for helpful discussions.
D.J.W.\ is grateful to the Royal Society for financial support.
The work of the FOM Institute is part of the
scientific program of FOM and is supported by the Nederlandse
Organisatie voor Wetenschappelijk Onderzoek (NWO).

\begin{table}
\begin{center}
\begin{minipage}{8.5cm}
\caption{\label{table:global} Energies of the lowest minima
found for \NaCl{N} clusters with the Coulomb plus Born-Meyer and
Welch potentials.}
\vglue0.1mm
\begin{tabular}{ccccc}
 & \multicolumn{2}{c}{C+BM} & \multicolumn{2}{c}{Welch} \\
 \cline{2-3}\cline{4-5}
\noalign{\vspace{1pt}}
 $N$ & Energy/eV & PG  & Energy/eV & PG \\
\hline
   1  &     $-7.712$        & $D_{\infty h}$ &      $-7.960$   & $D_{\infty h}$ \\
   2  &    $-14.948$        & $D_{\infty h}$ &     $-15.272$   & $C_{2v}$ \\
   3  &    $-22.452$        & $C_{3v}$ &           $-23.040$   & $C_{3v}$ \\
   4  &    $-30.204$        & $C_{4v}$ &           $-30.871$   & $C_{4v}$ \\
   5  &    $-37.386$        & $C_{2v}$ &           $-38.221$   & $C_1$ \\
   6  &    $-44.891$        & $C_s$ &              $-45.793$   & $C_s$ \\
   7  &    $-52.511$        & $C_{2v}$ &           $-53.456$   & $C_{2v}$ \\
   8  &    $-60.208$        & $C_{4v}$ &           $-61.163$   & $C_{4v}$ \\
   9  &    $-67.647$        & $C_{3h}$ &           $-68.979$   & $C_{3h}$ \\
  10  &    $-75.141$        & $C_s$ &              $-76.398$   & $C_2$ \\
  11  &    $-82.658$        & $C_s$ &              $-84.007$   & $C_1$ \\
  12  &    $-90.482$        & $C_{4v}$ &           $-91.668$   & $C_{4v}$ \\
  13  &    $-98.672$        & $O_h$ &              $-99.756$   & $O_h$ \\
  14  &   $-105.567$        & $C_s$ &             $-106.880$   & $C_s$ \\
  15  &   $-113.132$        & $C_1$ &             $-114.570$   & $C_s$ \\
  16  &   $-121.086$        & $C_{2v}$ &          $-122.497$   & $C_{2v}$ \\
  17  &   $-128.703$        & $C_{4v}$ &          $-129.994$   & $C_{4v}$ \\
  18  &   $-135.761$        & $C_1$ &             $-137.381$   & $C_1$ \\
  19  &   $-143.611$        & $C_s$ &             $-145.255$   & $C_2$ \\
  20  &   $-151.084$        & $C_s$ &             $-152.667$   & $C_s$ \\
  21  &   $-158.972$        & $C_{4v}$ &          $-160.448$   & $C_{4v}$ \\
  22  &   $-167.158$        & $D_{4h}$ &          $-168.576$   & $D_{4h}$ \\
  23  &   $-174.123$        & $C_s$ &             $-175.757$   & $C_1$ \\
  24  &   $-181.602$        & $C_1$ &             $-183.331$   & $D_3$ \\
  25  &   $-189.525$        & $C_s$ &             $-191.227$   & $C_s$ \\
  26  &   $-197.173$        & $C_{4v}$ &          $-198.766$   & $C_{4v}$ \\
  27  &   $-204.809$        & $C_{2v}$ &          $-206.602$   & $C_{2v}$ \\
  28  &   $-212.385$        & $C_s$ &             $-214.152$   & $C_s$ \\
  29  &   $-219.969$        & $C_s$ &             $-221.732$   & $C_1$ \\
  30  &   $-227.494$        & $C_s$ &             $-229.315$   & $C_s$ \\
  31  &   $-235.584$        & $D_{4h}$ &          $-237.305$   & $D_{4h}$ \\
  32  &   $-242.755$        & $C_s$ &             $-244.505$   & $C_s$ \\
  33  &   $-250.295$        & $C_1$ &             $-252.320$   & $C_1$ \\
  34  &   $-258.304$        & $C_s$ &             $-260.187$   & $C_s$ \\
  35  &   $-265.748$        & $C_s$ &             $-267.756$   & $C_s$ \\
\end{tabular}
\end{minipage}
\end{center}
\end{table}

\begin{table}
\begin{center}
\begin{minipage}{8.5cm}
\caption{\label{table:min}The fifteen lowest energy (NaCl)$_{35}$Cl$^-$ minima for the 
Coulomb plus Born-Meyer potential. The energies of the minima after reoptimization
with the Welch potential are also given.}
\begin{tabular}{ccccccc}
 & \multicolumn{2}{c}{Rank} & \multicolumn{2}{c}{Energy/eV} & PG & structure \\
 \cline{2-3}\cline{4-5}
\noalign{\vspace{1pt}}
 & C+BM & Welch & C+BM & Welch \\
\hline
 A & 1 & 1 & -265.748 & -267.756 & $C_s$ & $5 \times 5 \times 3$ \\
 B & 2 & 4 & -265.741 & -267.653 & $C_1$ & $5 \times 5 \times 3$ \\
 C & 3 & 2 & -265.739 & -267.721 & $C_1$ & $5 \times 5 \times 3$ \\
 D & 4 & 6 & -265.685 & -267.582 & $C_s$ & $6 \times 4 \times 3$ \\
 E & 5 & 3 & -265.658 & -267.656 & $C_1$ & $6 \times 4 \times 3$ \\
 F & 6 & 5 & -265.652 & -267.588 & $C_{2v}$ & $5 \times 5 \times 3$ \\
 G & 7 & 8 & -265.606 & -267.524 & $C_s$ & $6 \times 4 \times 3$ \\
 H & 8 & 7 & -265.601 & -267.573 & $C_1$ & $5 \times 5 \times 3$ \\
 I & 9 & 12 & -265.592 & -267.476 & $C_{4v}$ & $8 \times 3 \times 3$ \\
 J & 10 & 25 & -265.573 & -267.417 & $C_s$ & $6 \times 4 \times 3$ \\
 K & 11 & 13 & -265.557 & -267.467 & $C_2$ & $5 \times 5 \times 3$ \\
 L & 12 & 29 & -265.543 & -267.400 & $C_s$ & $5 \times 4 \times 4$ \\
 M & 13 & 15 & -265.542 & -267.465 & $C_s$ & $6 \times 4 \times 3$ \\
 N & 14 & 16 & -265.519 & -267.445 & $C_s$ & $5 \times 5 \times 3$ \\
 O & 15 & 9 & -265.507 & -267.502 & $C_1$ & $5 \times 4 \times 4$ \\
\end{tabular}
\end{minipage}
\end{center}
\end{table}

\begin{table}
\begin{center}
\begin{minipage}{8.5cm}
\caption{\label{table:paths}Details of the pathways 
between the lowest energy cuboidal nanocrystals. 
We give the energy difference between the highest energy transition state 
and the lowest energy minimum of the higher energy nanocrystal (the barrier), 
and the path length~\protect\cite{Wales94b}
for the paths that minimize these quantities.
The values of the barrier in brackets are for the Welch potential
and were obtained by reoptimization of a selection of
low energy pathways; an extensive search of the PES
was not conducted with the latter potential.
For the $5\times 4\times 4$ nanocrystal we give the barriers for the minima L and O 
which are the lowest energy minima of this type for the Coulomb plus Born-Meyer and the
Welch potentials, respectively.}
\vglue0.1mm
\begin{tabular}{llcc}
 \hfil From & \hfil To & barrier/eV & path length/\AA \\
\hline
 $6\times 4\times 3$ & $5\times 5\times 3$ & 0.78 (0.69) & 33.5 \\
 $8\times 3\times 3$ & $5\times 5\times 3$ & 1.06 (0.91) & 69.8 \\
 $5\times 4\times 4$ L & $5\times 5\times 3$ & 0.62 (0.41)& 66.3 \\
 $5\times 4\times 4$ O & $5\times 5\times 3$ & 0.60 (0.54) & 46.6 \\
\noalign{\vspace{1pt}}
 $8\times 3\times 3$ & $6\times 4\times 3$ & 1.06 (0.91) & 36.3 \\
 $5\times 4\times 4$ L & $6\times 4\times 3$ & 0.63 (0.44) & 57.1 \\
 $5\times 4\times 4$ O & $6\times 4\times 3$ & 0.56 (0.54) & 14.4 \\
\noalign{\vspace{1pt}}
 $5\times 4\times 4$ L & $8\times 3\times 3$ & 1.01 (0.83) & 92.6 \\
 $5\times 4\times 4$ O & $8\times 3\times 3$ & 0.97 (0.91) & 50.7 \\
\noalign{\vspace{1pt}}
 $5\times 4\times 4$ O & $5\times 4\times 4$ L & 0.60 (0.43) & 42.8 \\
\end{tabular}
\end{minipage}
\end{center}
\end{table}

\end{multicols}
\end{document}